# Quasi-monoenergetic electron beams from laser plasma acceleration using nitrogen gas jet

B. S. Rao[1], A. Moorti[1,2], J. A. Chakera[1,2], P. A. Naik[1,2], and P. D. Gupta[2]

[1]Laser Plasma Division and [2]Homi Bhabha National Institute, Raja Ramanna Centre for Advanced Technology, Indore-452013, India

E-mail: sunnybobbili.raobs@gmail.com

## ABSTRACT

An experimental investigation on laser plasma acceleration of electrons has been carried out using 3 TW, 45 fs duration titanium sapphire laser pulse interaction with nitrogen gas jet at intensity of $2 \times 10^{18}$ W/cm$^2$. We have observed stable generation of well collimated electron beam with divergence and pointing variation ~ 10 mrad from the nitrogen gas jet plasma at an optimum plasma density around $3 \times 10^{19}$ cm$^{-3}$. The energy spectrum of the electron beam was quasi-monoenergetic with average peak energy and charge around 25 MeV and 30 pC respectively. The results will be useful for better understanding and control of ionization injection and laser wakefield acceleration of electrons in high-Z gases and also to develop practical laser wakefield accelerators for various applications including injectors for high energy accelerators.

# INTRODUCTION

Laser wakefield acceleration (LWFA) of electrons has attracted a lot of interest since its proposal in 1979 as this technique offers high accelerating longitudinal electric field ~ 100 GV/m [1,2]. Presently, LWFA is considered to be a promising technique to develop compact next generation TeV electron-positron collider which can be potentially less expensive [3]. Further, the electron beams produced from LWFA have unique characteristics viz., bunch length ~ few fs, peak current ~ few kA, and emittance ~ $1\pi$ mm.mrad [4, 5]. These properties make the LWFA attractive also for developing compact high brightness ultra-short pulsed radiation sources including high gain short wavelength free electron lasers [6]. There has been tremendous progress in this field particularly in recent years owing to continuous efforts to push the limits of achievable energy and at the same time to improve the quality and stability of the generated electron beam. Generation of multi-GeV electron beams from LWFA have been demonstrated recently using peta-watt class lasers and cm-scale plasma produced in gas jets [7], gas cell [8], and capillary discharge targets [9]. Generation of moderately stable and quasi-monoenergetic electron beams up to sub-GeV energy has also been demonstrated by controlled injection of electrons by various techniques [10-16].

In a typical LWFA, an ultra-short intense laser pulse focused in a gas medium and the gas atoms or molecules are rapidly ionized by the foot of the laser pulse. The highly intense part of the laser pulse interacts with the plasma produced and expels the plasma electrons radially outwards due to ponderomotive force and forms nearly spherical shaped plasma bubble / cavity consisting dense sheath of electrons at its edges. The plasma ions do not respond to the ponderomotive force on the time scale fs laser pulse duration and therefore form immobile background inside the plasma bubble. The electric field inside the bubble due to charge separation could be very high of the order of ~ 100 GV/cm. The bubble propagates along with the laser pulse with velocity close to velocity of light in vacuum (*c*). The bubble formation is efficient when the laser pulse parameters i.e., focal spot ($w_0$) and pulse length ($c\tau_L$, where $\tau_L$ is the pulse duration) are on the order of plasma wavelength ($\lambda_p$). When the driver laser pulse is intense enough, it can lead to self-injection of electrons in to the bubble [17]. The injected electrons can be trapped and accelerated by the strong longitudinal electric field of the bubble resulting in generation of energetic electron beam. Low-Z gases like hydrogen ($H_2$) / helium (He) are mostly used as target for LWFA experiments to minimize/avoid defocusing effects due to laser induced ionization and also due to simplicity of the interaction process. However, use of $H_2$ poses hazard as it is highly reactive and explosive. Helium, on the other hand, is inert and safe to use but it is very expensive as it is a rare gas. High-Z gases for LWFA are sparsely explored till now. Particularly, despite $N_2$ being readily available and cheap due to its abundance in ambient air it was not explored enough for LWFA. Recently, investigation of high-Z gases [18-20], particularly $N_2$ for LWFA has gained renewed interest due to the positive role of "ionization induced injection" [21, 22]. Experiments and simulations show that ionization induced injection occurs at lower laser intensity of $a_0$ ~2, { $a_0$=0.86×[$I_L$($10^{18}$ W/cm$^2$)×$\lambda^2$(μm)]$^{0.5}$, with $I_L$ intensity and λ wavelength of the laser} compared to self-injection which requires $a_0 >$ 4.

In this paper, we present an experimental investigation on laser plasma acceleration of electrons in pure nitrogen gas jet plasma using 3 TW, 45 fs duration Ti:Sapphire laser focused to intensity 2×$10^{18}$ W/cm$^2$. Relativistic self-guiding of the laser pulse and generation of quasi-monoenergetic electron beam with divergence ~10 mrad, energy ~ 25 MeV and charge ~ 30 pC was observed at an optimum plasma density. The results will be useful for better understanding of ionization injection and laser wakefield acceleration of electrons in high-Z gases. Particularly, the quasi-monoenergetic electron beams produced from the nitrogen gas jet may be used to inject in to a much higher energy laser plasma accelerator. The use of nitrogen gas for LWFA provides additional control on electron beam parameters through K-shell ionization process. In addition, the nitrogen gas is available at much lower cost and safer to use when compared to widely used helium / hydrogen gas jet.

**EXPERIMENTAL SET-UP**

Figure 1 shows a schematic of the experimental setup used for the laser plasma acceleration. The experiment was performed using the titanium sapphire laser system at Raja Ramanna Centre for Advanced Technology (RRCAT), Indore, India. The laser system provides horizontally polarized laser pulses of FWHM duration, $\tau_L \gtrsim 45\ fs$ with central wavelength, $\lambda_0 \simeq 800$ nm and bandwidth, $\Delta\lambda \simeq 20$ nm. The laser beam was focused by an f/5 gold coated off-axis parabola along 1.2 mm width of a rectangular (1.2 mm x10 mm) supersonic nozzle at a height of 1 mm above the nozzle orifice. The full width at $1/e^2$ maximum size of the focal spot is 27 μm × 15 μm and is approximated with an equivalent area circular spot of $1/e^2$ width, $2w_0 = 20$ μm to estimate the peak intensity. Peak power of the laser pulse on the target is $P_L = 0.92E_L/\tau_L \simeq 3$ TW for the laser pulse energy, $E_L$=145 mJ contained in the focal spot and assuming Gaussian temporal profile. The peak intensity in the focal spot is estimated to be $I_L=2P_L/\pi w_0^2 \simeq 2\times10^{18}$ W/cm$^2$ ($a_0 \simeq 1$). The gas jet was well characterized for number density of gas atoms at different backing pressures using interferometry measurements [23]. The focused laser pulse when interacts with the nitrogen gas, the nitrogen atoms will be tunnel ionized to N$^{5+}$ state by the foot of the laser pulse at intensity ~ $1\times10^{16}$ W/cm$^2$ and further ionization to N$^{6+}$ requires intensity ~ $1\times10^{19}$ W/cm$^2$ which is 5 times the peak intensity of the laser pulse at focus in vacuum. We estimate the plasma density in the interaction region accordingly from the known number density of nitrogen atoms for a given backing pressure of the gas jet. A single-shot electron spectrograph consisting of a circular permanent magnet poles with 50 mm diameter, 9 mm pole gap, and effective magnetic field $B_{eff}$ = 0.46 T to deflect the electrons, and a DRZ-High ($Gd_2O_2S$:Tb) phosphor screen imaged with a 16-bit CCD camera was used for electron detection and measuring the energy spectrum. A slit was used at the entrance of the magnet to allow electrons within 8 mrad acceptance angle in the plane of electron dispersion. The low energy cut-off of the electron spectrograph was 10 MeV. The electron beam charge was estimated using the absolute calibration data of the phosphor screen [24]. The electron beam profile and pointing variation was measured by displacing the magnet away from electron beam path. The laser-plasma interaction was imaged from the non-linear Thomson side-scattering radiation at second harmonic (400 nm) of the laser, with 5× magnification using a 12-bit CCD camera. A narrow band-pass filter with transmission wavelength at 400 ± 20 nm was kept in front of the CCD for collecting second harmonic radiation. A small fraction of the drive laser beam with variable delay relative to the main laser beam was used as a probe beam (back-lighter) to record the shadow gram of the interaction region with 2 ps resolution. The probe beam was sent perpendicular to the incident laser axis and the gas jet flow direction and a highly reflecting multilayer coated mirror for 800 nm radiation was placed in front of the side imaging CCD to reflect the probe beam on to a separate 12-bit CCD while transmitting the 400 nm radiation to reach the side imaging CCD, as depicted in Fig. 1. A narrow band pass filter that allows the wavelengths within 800 ± 20 nm was kept in front of the shadow gram imaging CCD.

## RESULTS AND DISCUSSIONS

We have scanned the plasma density by adjusting the backing pressure of the gas jet and also fine tuned the laser focus position in the gas jet until we observe Thomson side scattering images of laser self-guiding and also generation of intense electron beam from laser plasma acceleration. A well defined and low divergence electron beam (Fig. 2a) was observed at gas pressure of about 10 bar with corresponding estimated plasma density of $n_e \sim 3\times10^{19}$ cm$^{-3}$ (assuming laser induced ionization up to N$^{5+}$). The well-defined low divergence electron beam also has low intensity diffused background halo (see the background in Fig. 2a) of electrons surrounding it. At plasma density of $n_e \sim 3\times10^{19}$ cm$^{-3}$, the Thomson scattering images showed laser self-guiding channels of length of 440±50 μm. Although self-guiding images were also observed at relatively higher gas pressure, the produced electron beam has larger divergence and poor stability. Interestingly, the low divergence electron beam shown in Fig. 2a was highly reproducible with >90% probability in series of about 10 – 20 shots in most of the experimental runs. The transverse intensity profile of the electron beam is elliptical in shape with its major axis along the horizontal direction which is along the laser polarization axis. The ellipticity (defined here as the ratio of the length of major axis to minor axis) of the electron beam was estimated to be $\varepsilon = 1.9^{+0.9}_{-0.6}$. The FWHM divergence of the electron beam is $\Delta\theta_X = 14.5^{+5.3}_{-4.1}$ mrad and $\Delta\theta_Y = 8.0^{+2.0}_{-2.3}$ mrad in the horizontal and vertical direction respectively. The values of root-mean-square (RMS) deviations from the average value are indicated as subscript and superscript to the average value. The RMS value was independently calculated for the occurrences with higher-than-average and lower-than-average value to account for asymmetric distributions. The pointing variation of the low divergence electron beam produced is shown in Fig. 2b. The RMS pointing deviation from the mean pointing angle of the electron beam in horizontal and vertical direction was <$\theta_X$> = 11.9 mrad and <$\theta_Y$> = 8.1 mrad respectively. It is interesting to note that the pointing variation of the electron beam in the horizontal and vertical directions is close to the divergence in respective directions. The charge contained in the electron beam was $\sim 27.9^{+38.0}_{-7.6}$ pC. The reproducibility and pointing variation of the electron beam were observed to degrade quickly with increasing plasma densities although overall charge of the electron beam increased.

A series of 15 images of electron energy spectra recorded at nitrogen plasma density $\simeq 3\times10^{19}$ cm$^{-3}$ in one of the experimental runs is shown in Fig. 3. The electron beam is quasi-monoenergetic with peak energy, $E_{peak} = 25.5^{+5.3}_{-7.9}$ MeV with relative FWHM energy spread ($\Delta E_{FWHM}/E_{peak}$) of $28.3^{+13.7}_{-13.4}$ %. There will be some uncertainty in energy due to the pointing variation of the electron beam in the dispersion plane. Since the acceptance angle (8 mrad) of the slit in front of the spectrograph was smaller than the pointing variation of the beam in the dispersion plane, actual value of the uncertainty is decided by the acceptance angle of the spectrograph. The uncertainty in the peak energy due to the pointing variation of the electron beam is about 20% at 25 MeV and 28% at 35 MeV. The larger divergence and the larger pointing variation w.r.t. the angular acceptance of the spectrograph also meant that a fractional cross-section of the electron beam has been intercepted by the edge(s) of the slit. Therefore, it may be inferred that the actual charge of the electron beam is higher and the shot-to-shot charge variation is lower than that reported here. In addition, since the maximum angular dimension of the beam on the phosphor screen in the case of mono-energetic electrons would be ~8 mrad. This finite size of the beam in the dispersion plane causes an apparent energy spread of 20.3% around average peak energy of about 25 MeV. A histogram of the peak energy distribution of the electron beam is shown in Fig. 4. Although about 50% of the spectra have low energy background characteristic of ionization induced injection the rest of the spectra showed quasi-monoenergetic beams free from low-energy background The spectral intensity profile of one such electron beam (in Shot no. 4 in Fig. 3) is shown in Fig. 5.

The results can be understood invoking the role of ionization induced injection [21, 22]. Earlier, we have investigated LWFA in pure He gas jet under same laser interaction parameters and observed self-injection of electrons at higher gas pressure (32 bar) or plasma density ($6\times10^{19}$ cm$^{-3}$) [16]. However, in the present experiment with the pure nitrogen gas the self-injected electrons are observed at lower density of $3\times10^{19}$ cm$^{-3}$ (gas pressure ~ 10 bar) which clearly suggests the role of ionization induced injection. The lower threshold density for electron self-injection in the case of N$_2$ could be due to injection of electrons released by the ionization of N$^{5+}$ inside the plasma bubble. Although the initial laser intensity $2\times10^{18}$ W/cm$^2$ in vacuum is not sufficient for ionization of N$^{5+}$, the laser pulse evolution through self-focusing and self-compression over the initial interaction length in the plasma may enhance the laser intensity to ~ $1\times10^{19}$ W/cm$^2$ ($a_0$ ~ 2) required for ionization of N$^{5+}$. This could happen before the laser pulse intensity (or $a_0 > 4$) reaches the threshold for transverse self-injection by trajectory crossing at vertex of the "bubble". The maximum energy achievable in bubble regime is given by $W_{gain} = (2/3)(n_c/n_e)a_0 m_0 c^2$, where $n_c$ or $n_e$ is the critical or plasma density and $m_0$ is the rest mass of electron [17]. Assuming $a_0$ ~ 2, the maximum energy expected in our experiment will be about 38 MeV which is the maximum energy observed in the present experiment [see Fig. 5]. The lower mean peak energy of 25.5 MeV observed could be due to the beam loading effect [25] caused by higher beam charge in low energy beams. The larger relative energy spread is again characteristic of ionization injection which occurs for longer duration in the bubble. However, ionization injection can be self-terminated under certain conditions [26] resulting in relatively low charge but less energy spread and high energy electron beams as also observed in present experiments (Fig. 5).

In the present experiment, the laser pulse evolves through its interaction with the nitrogen plasma and the electron injection takes place only when the laser attains peak intensity enough for laser induced ionization of background $N^{5+}$ ions. The fact that we did not observe electron beam on phosphor screen at lower plasma density (<10 bar) indicates that the laser pulse could not evolve sufficiently at the corresponding plasma density for causing ionization injection. Further, the energy of the electron beam at relatively higher gas pressure or plasma density was observed to have relatively lower energy electrons, with broad spectrum. Ionization of $N^{5+}$ at higher gas pressure is expected to inject more electrons into the bubble resulting in generation of higher charge electrons but with large divergence and lower energy due to enhanced beam loading effect. Therefore, our observations of higher charge, higher divergence, and low energy electron beam at higher gas pressure further signifies the role of ionization injection in the experiment.

## SUMMARY

In summary, an experimental study on laser driven plasma based electron acceleration was performed in pure nitrogen gas jet using 45fs, 3 TW titanium sapphire laser. We found that low divergence ~ 10 mrad, quasi-monoenergetic electron beams of about 25 MeV energy and 30 pC charge could be produced from the gas-jet target. The results will be useful in the efforts towards the development of controlled practical laser wakefield accelerators for various applications in future including as injectors for staged LWFA.

## ACKNOWLEDGMENTS

The authors acknowledge the laser and other technical support by R. A. Khan, R. A. Joshi, and R. K. Bhat, and mechanical support by R. P. Kushwaha, S. Sebastin, and R. Parmar during the experiment.

**FIGURES**

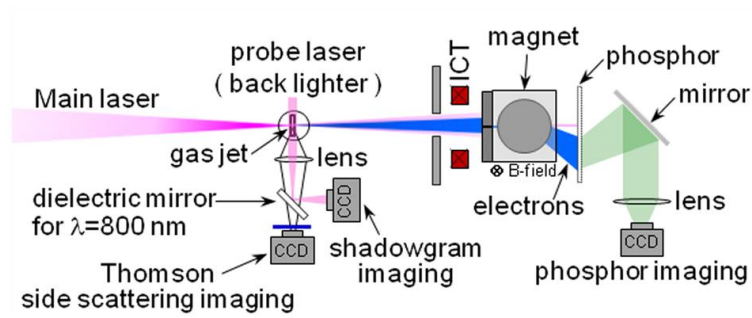

Fig.1. Schematic of the experimental set-up used for laser plasma acceleration.

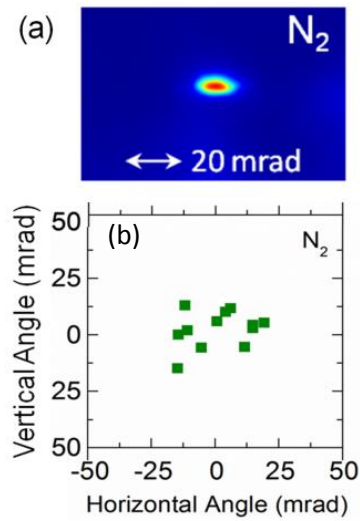

Fig. 2. (a) Image of the electron beam produced from nitrogen gas jet at plasma density of ~$3\times10^{19}$ cm$^{-3}$. (b) Pointing variation of the electron beam.

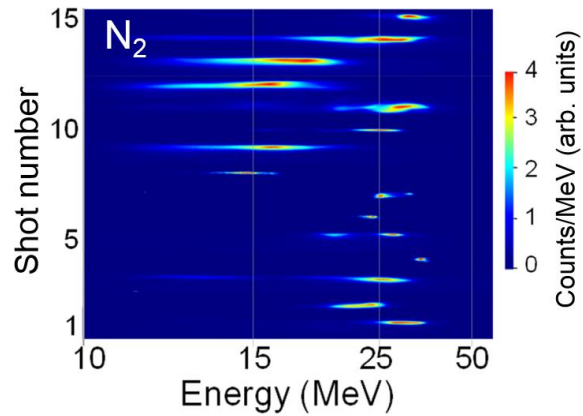

Fig. 3. Images of energy dispersed electron beams recorded in 15 consecutive laser shots at nitrogen plasma density of ~$3\times10^{19}$ cm$^{-3}$. The color scale is normalized for each shot.

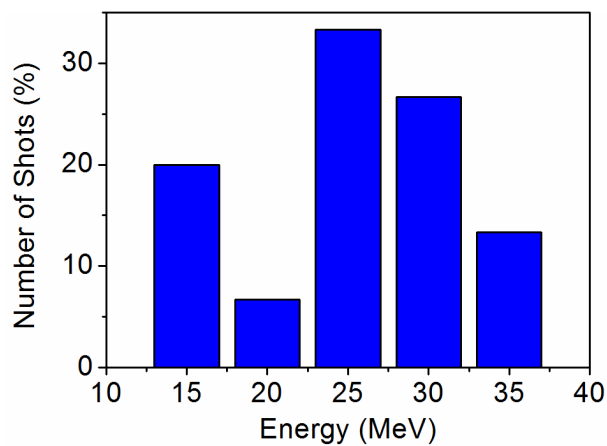

Fig. 4. A histogram of the peak energy of the quasi-monoenergetic electron beams produced from nitrogen gas jet at plasma electron density of ∼3 ×$10^{19}$ cm$^{-3}$. The histogram shows 5 bins, each of 5 MeV width.

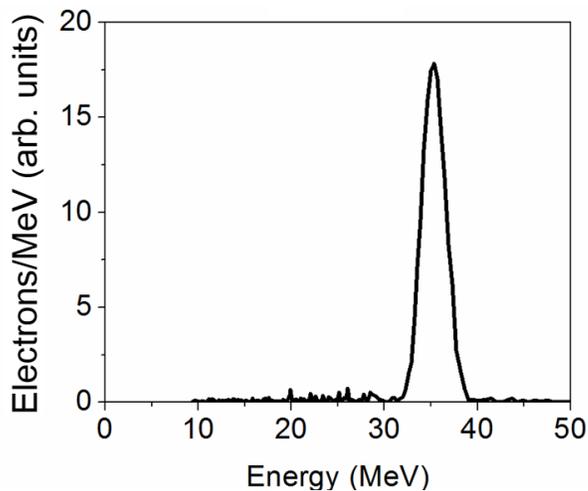

Fig.5. Energy spectrum of a narrow energy spread quasi-monoenergetic electron beam (corresponding to shot no.4 in Fig.3) with almost no low energy background produced from nitrogen gas jet at plasma density of ~3×$10^{19}$ cm$^{-3}$.